\documentclass[epjH]{svjour}
\usepackage{graphics}
\begin{document}
\title{Nationalism and internationalism in science: The case of the discovery of cosmic rays}
\author{Per Carlson\inst{1}
\and Alessandro De Angelis\inst{2}
}
\institute{KTH and the Oskar Klein Centre, Stockholm, Sweden \and Max-Planck-Institut f\"ur Physik
(Werner-Heisenberg-Institut),
M\"unchen, Germany\thanks{On leave of absence from Universit\`a di Udine, Italy.}; INFN and INAF Trieste, Italy; LIP/IST, Lisboa, Portugal }
\abstract{The discovery of cosmic rays, a milestone in science, comprised of scientists in Europe and the New World and took place during a period characterised by nationalism and lack of communication. Many scientists that took part in this research a century ago were intrigued by the penetrating radiation and tried to understand the origin of it. Several important contributions to the discovery of the origin of cosmic rays have been forgotten in particular that of Domenico Pacini, who in June 1911 demonstrated by studying the decrease of radioactivity with an electroscope immersed in water that cosmic rays could not come from the crust of the Earth. Several historical, political and personal facts might have contributed to the substantial disappearance of Pacini from the history of science.
} 
\maketitle
\section{Introduction}

By 1785 de Coulomb found \cite{Cou1785} that electroscopes can spontaneously discharge due to the action of the air and not because of defective insulation. After dedicated studies by Faraday around 1835 \cite{Far1835}, Crookes observed in 1879 \cite{Cro1879} that the speed of discharge of an electroscope decreased when pressure was reduced. The explanation of this phenomenon came in the beginning of the 20th century and paved the way to one of mankind's revolutionary scientific discoveries: cosmic rays. 
Several  reviews have been written on the history of the research on cosmic rays
 \cite{montgo,janossy,leprince,rossi,hillas,wilson,wigandr,millikanr,comptonr,swann,xu,ginz,puppi}.

\section{The puzzle of atmospheric ionisation}

Spontaneous radioactivity was discovered in 1896 by Becquerel \cite{Beq1896}. A few years later, Marie and Pierre Curie \cite{Cur1898} discovered that the elements Polonium and Radium suffered transmutations generating radioactivity: such transmutation processes were then called ``radioactive decays''. In the presence of a radioactive material, a charged electroscope promptly discharges. It was concluded that some elements were able to emit charged particles, that in turn caused discharge of electroscopes.  The discharge rate was then used to gauge the level of radioactivity. 

Following the discovery of radioactivity, a new era of research in discharge physics was then opened, this period being strongly influenced by the discoveries of the electron and of positive ions. During the first decade of the 20th century results on the study of ionisation phenomena came from several researchers in Europe and in the New World. 

Around 1900, Wilson \cite{Wil1901} and Elster and Geitel \cite{ElG1900} improved the technique for the careful insulation of electroscopes in a closed vessel, thus improving the sensitivity of the electroscope itself. As a result, they could make quantitative measurements of the rate of spontaneous discharge. They concluded that such a discharge was due to ionising agents coming from outside the vessel. The obvious questions concerned the nature of such radiation, and whether it was of terrestrial or extra-terrestrial origin. The simplest hypothesis was that its origin was related to radioactive materials, hence terrestrial origin was a commonplace assumption. An experimental confirmation, however, seemed hard to achieve. Wilson \cite{Wil1901} tentatively made the visionary suggestion that the origin of such ionisation could be an extremely penetrating extra-terrestrial radiation. However, his investigations in tunnels with solid rock overhead showed no reduction in ionisation \cite{Wil1901} and could therefore not support an extra-terrestrial origin. An extra-terrestrial origin, though now and then discussed (for example by Richardson \cite{Ric1906}), was dropped for many years.

In 1903 Rutherford and Cooke \cite{RuC1903} and also McLennan and Burton  \cite{MLB1903} showed that the ionisation was significantly reduced when the closed vessel was surrounded by shields of metal kept free from radioactive impurity. This showed that part of the radiation came from outside. Later investigations showed that the ionisation in a closed vessel was due to ``penetrating radiation'' partly from the walls of the vessel and partly from outside. 

The situation in 1909 is well summarised by Kurz \cite{kurz}  and by Cline \cite{Cline}.  The spontaneous discharge was consistent with the hypothesis that even in insulated environments background radiation did exist. In the 1909 review by Kurz  three possible sources for the penetrating radiation are discussed: an extra-terrestrial radiation possibly from the Sun \cite{Ric1906}, radioactivity from the crust of the Earth, and radioactivity in the atmosphere. 
Kurz concludes from the ionisation measurements in the lower part of the atmosphere that an extra-terrestrial radiation was unlikely.
It was generally assumed that large part of the radiation came from radioactive material in the crust. Calculations were made of how the radiation should decrease with height (see for example Eve \cite{eve}) and measurements were performed.

\begin{figure}
\begin{center}
\resizebox{0.7\columnwidth}{!}{\includegraphics{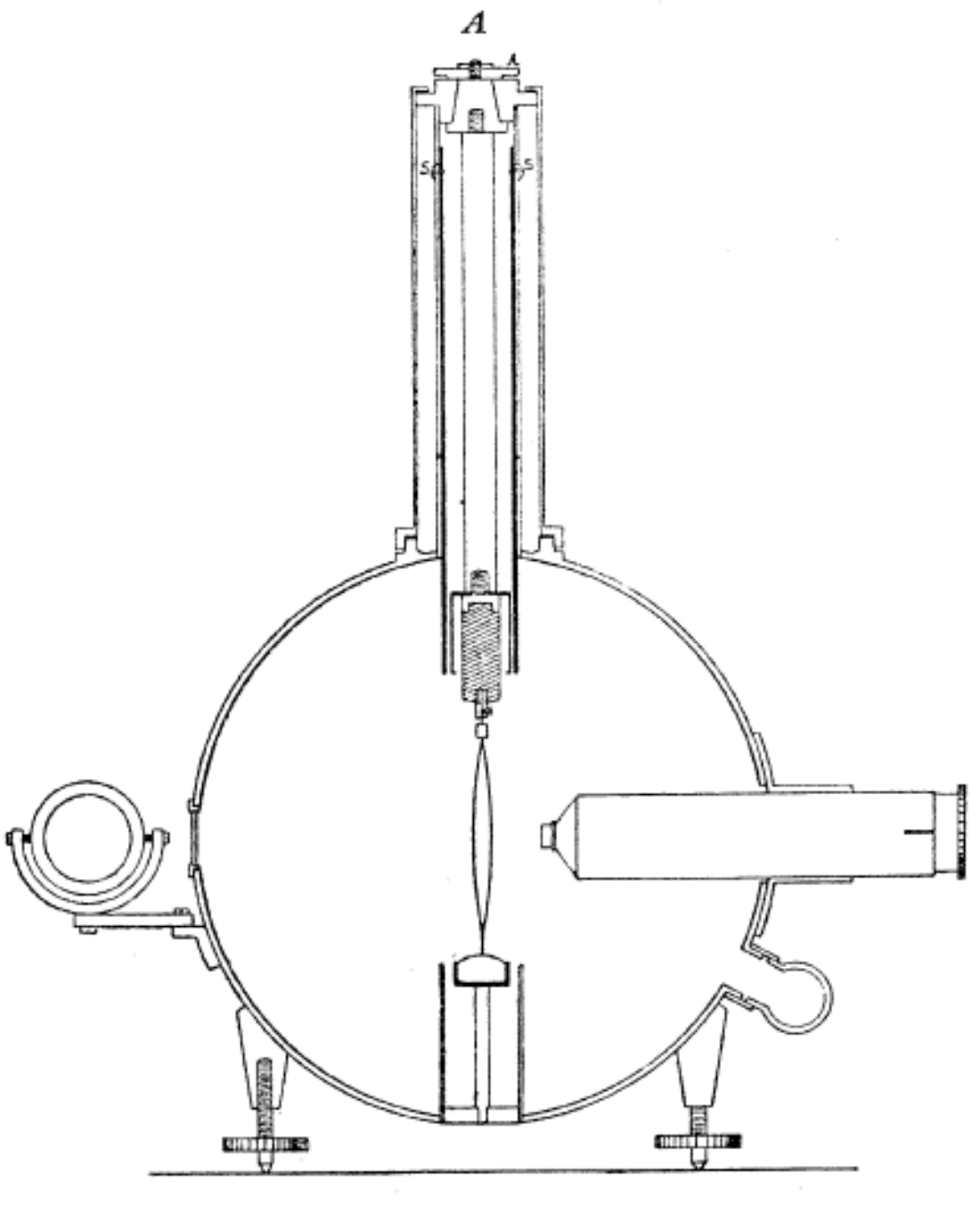} }
\end{center}
\caption{Fig. 1. The Wulf electroscope \cite{Wul1910}. The 17 cm diameter cylinder with depth 13 cm was made
of Zinc. To the right is the microscope that measured the distance between the two silicon
glass wires illuminated using the mirror to the left. The air was kept dry using Sodium in the small container below the microscope. According to Wulf \cite{Wul1910}, with 1.6 ion pairs
per second produced, the tension was reduced by 1 volt, the sensitivity of the instrument,
as measured by the decrease of the inter-wire distance.}
\label{fig:1}       
\end{figure}

Father Theodor Wulf, a German scientist and a Jesuit priest serving in the Netherlands and then in Rome, had the idea to check the variation of radioactivity with height to test its origin. In 1909 \cite{Wul1910} Wulf, using an improved  electroscope (Fig. \ref{fig:1}) in which the two leaves had been replaced by two metalised silicon glass wires, with a tension spring made also by glass in between, measured the rate of ionisation at the top of the Eiffel Tower in Paris (300 m above ground). Supporting the hypothesis of the terrestrial origin of most of the radiation, he expected to find at the top less ionisation than on the ground. The rate of ionisation showed, however, too small a decrease to confirm the hypothesis. He concluded that, in comparison with the values on the ground, the intensity of radiation ``decreases at nearly 300 m [altitude] was not even to half of its ground value''; while with the assumption that radiation emerges from the ground there would remain at the top of the tower ``just a few percent of the ground radiation'' \cite{Wul1910}. Wulf's observations were of great value, because he could take data at different hours of the day and for many days at the same place. For a long time, Wulf's data were considered as the most reliable source of information on the altitude effect in the penetrating radiation. However Wulf concluded that the most likely explanation of his puzzling result was still emission from the soil.

In 1909 by balloon ascent Bergwitz found \cite{Bergwitz} that the ionisation at 1300 m altitude had decreased to about 24\% of the value at ground, consistent with expectations if the radiation came from the Earth's surface. However, Bergwitz results were questioned because the electrometer was damaged during the flight (see, for example \cite{Goc1911}). Later Bergwitz investigated electrometers \cite{ber1913} and in 1915 reported results from a comparison of the ionisation on ground and at 80 m altitude \cite{ber1915}. No significant decrease of the ionisation was observed. Other measurements with similar results were also made (McLennan and Macallum \cite{McLennan}, Gockel \cite{Goc1909}). The general interpretation was that radioactivity was mostly coming from the Earth's surface  \cite{kurz,Cline}.

\subsection{Pacini and the measurements of attenuation in water}

The conclusion that radioactivity was mostly coming from the Earth's crust was questioned by the Italian physicist Domenico Pacini, who compared the rate of ionisation on mountains, over a lake, and over the sea \cite{Pac1909,Pac1910a,Pac1910b}. In 1911, he conducted a number of important experiments by immersing an electroscope deep in the sea \cite{Pac1912}. 

\begin{figure}
\begin{center}
\resizebox{0.5\columnwidth}{!}{\includegraphics{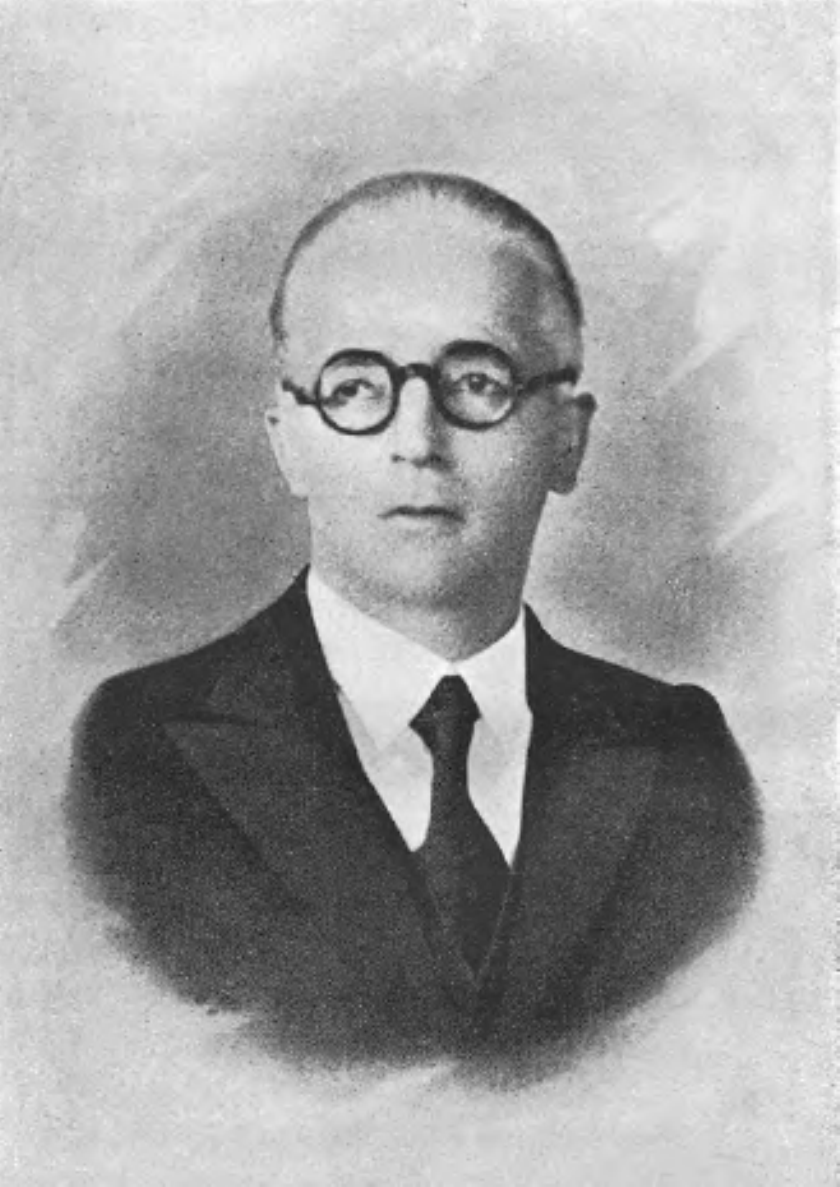} }
\end{center}
\caption{Domenico Pacini (from \cite{Riz1934a}; courtesy of the Societ\`a Italiana di Fisica).}
\label{fig:2}       
\end{figure}

Pacini (Fig. \ref{fig:2}) was born in 1878, in Marino, near Rome\footnote{The biographical information on Domenico Pacini is taken from the obituary \cite{Riz1934a,Riz1934b}, as most of the information on his work;
the documents related to his curriculum vitae for the applications for the ``libera docenza'' and the professorship were uncovered by \cite{rob} at the
Biblioteca Nazionale in Rome.}. He graduated in Physics in 1902 at the Faculty of Sciences of Rome  University. There, for the next three years, he worked as an assistant to Professor Blaserna studying electric conductivity in gaseous media. In 1906 Pacini was appointed assistant at Italy's Central Bureau of Meteorology and Geodynamics, heading the department that was in charge of studying thunderstorms and electric phenomena in the atmosphere. Pacini held the position until 1927, when he was promoted to Principal Geophysicist. Finally in 1928 he was appointed full professor of Experimental Physics at the University of Bari. Pacini died of pneumonia in Rome  in 1934, shortly after his marriage.

\begin{figure}
\begin{center}
\resizebox{0.75\columnwidth}{!}{ \includegraphics{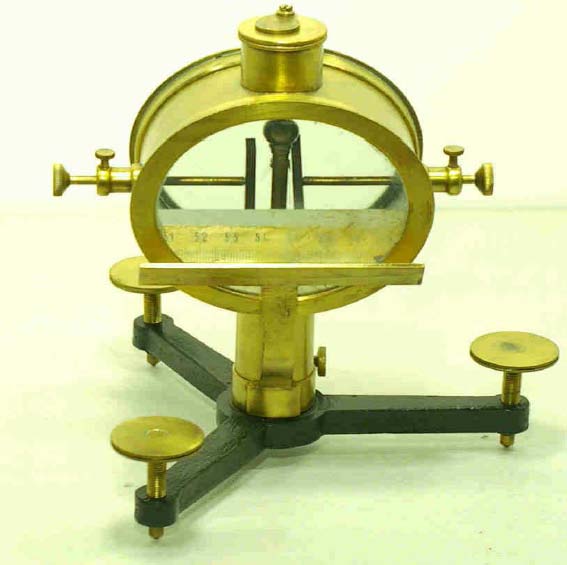} }
\end{center}
\caption{An Ebert-type electroscope (Elster \& Geitel) of the beginning of 20th century. This instrument consists of an electroscope to which a condenser (not visible in the photo) is attached as an integral part, and through this a known flux of air could be pumped in.
Courtesy of 
the Gabinetto di Fisica dellÕ Istituto Calasanzio di Empoli.}
\label{fig:3}       
\end{figure}

Pacini's important results on the penetrating radiation started with studies on electric conductivity in gaseous media performed at the University of Rome  during the early years of the 20th century.  During 1907-1912, he performed several detailed measurements on the conductivity of air, using an Ebert electroscope (Fig. \ref{fig:3}) to enhance the sensitivity (he could reach a sensitivity of one third of a volt). 

 In the first period Pacini made several measurements to establish the variations of the electroscope's discharge rate as a function of the environment. First he placed the electroscope on the ground and on the sea a few kilometres off the coast; the results were comparable. A summary of these results indicate, according to Pacini's conclusions, that ``in the hypothesis that the origin of penetrating radiations is in the soil, since one must admit that they are emitted at an almost constant rate (at least when the soil is not covered by remaining precipitations), it is not possible to explain the results obtained'' \cite{Pac1909}. 
In \cite{Pac1909} Pacini was the first to state that the results of many experiments on radiation could not be explained by radioactivity in the Earth's crust, a statement that was repeated by Gockel in \cite{Goc1909}.

\begin{figure}
\begin{center}
\resizebox{0.8\columnwidth}{!}{\includegraphics{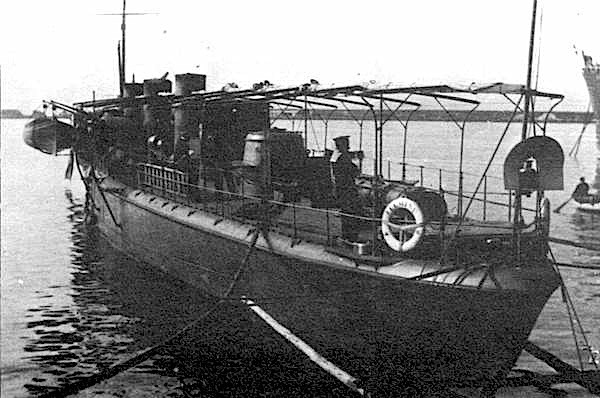} }
\end{center}
\caption{The cacciatorpediniere ``Fulmine'', used by Pacini for his measurements on the sea (courtesy of the Marina Militare Italiana).}
\label{fig:fulmine}       
\end{figure}

Pacini continued the investigations of radiation and developed in 1911 an experimental technique for underwater measurements. He found a significant decrease in the discharge rate when the electroscope was placed underwater. Pacini made his measurements over the sea in the Gulf of Genova, on an Italian Navy ship, the {\em cacciatorpediniere} (destroyer) ``Fulmine" (Fig. \ref{fig:fulmine}) from the Accademia Navale di Livorno.

He reported these measurements, the ensuing results, and their interpretation, in a note titled {\em La radiazione penetrante alla superficie ed in seno alle acque (Penetrating radiation at the surface of and in water)} \cite{Pac1912}. In that paper Pacini wrote: ``Observations carried out on the sea during the year 1910 \cite{Pac1910a} led me to conclude that a significant proportion of the pervasive radiation that is found in air had an origin that was independent of direct action of the active substances in the upper layers of the Earth's surface. ... [To prove this conclusion] the apparatus ... was enclosed in a copper box so that it could immerse in depth. ... From June 24 to June 31 [sic!] [1911], observations were performed with the instrument at the surface, and with the instrument immersed in water, at a depth of 3 metres.''  

With the apparatus at the surface 300 m from land, Pacini measured seven times during three hours
the discharge of the electroscope, obtaining a loss of 12.6 V/hour, corresponding to
11.0 ions per second (with a RMS of 0.5 V/hour); with the apparatus at a 3 m depth in the 7 m deep sea, he measured an average loss of 10.3 V per hour, corresponding to 8.9 ions per second
(with a RMS of 0.2 V/h). Consistent results were obtained during measurements at
the Lake of Bracciano.

The underwater measurement was 20\% lower than at the surface, consistent with absorption by water of a radiation coming from outside. ``With an absorption coefficient of 0.034 for water, it is easy to deduce from the known equation $I/I_0$ = exp(-d/$\lambda$), where d is the thickness of the matter crossed, that, in the conditions of my experiments, the activities of the sea-bed and of the surface were both negligible. É The explanation appears to be that, due to the absorbing power of water and the minimum amount of radioactive substances in the sea, absorption of radiation coming from the outside happens indeed, when the apparatus is immersed.'' Pacini concluded \cite{Pac1912}: ``[It] appears from the results of the work described in this Note that a sizable cause of ionisation exists in the atmosphere, originating from penetrating radiation, independent of the direct action of radioactive substances in the soil." As a curiosity, in 1910 Pacini looked for a possible increase in radioactivity during a passage of the Halley's comet (and he found no effect of the comet itself).

We note that a measurement of  ionisation  had been reported 1903 by McLennan and Burton \cite{MLB1903}, where
 the electroscope was placed in a container which in turn was closed in an iron tank, sometimes filled with 120 cm of water, sometimes shielded by metal layers. These authors observed a ionisation reduction up to 40\%. Their conclusion was in line with radioactivity in the air from known radioactive sources: ``From these results it is evident that the ordinary air of a room is traversed by an exceedingly penetrating radiation such as that which Rutherford has shown to be emitted by thorium, radium and the excited radioactivity produced by thorium and radium'' \cite{MLB1903}.

\subsection{Hess and the balloon measurements}
The need for balloon experiments became evident to clarify Wulf's observations on the effect of altitude (at that time and since 1885, balloon experiments were anyway widely used for studies of the atmospheric electricity). The first balloon flight had the purpose of studying the properties of penetrating radiation and was arranged in Switzerland in December 1909 with a balloon called Gotthard from the Swiss aeroclub. Alfred Gockel, professor at the University of Fribourg, ascended in a balloon to 4500 m above sea level (a.s.l.) during three successive flights, found \cite{Goc1910,Goc1911} that ioniation did not decrease with height as would be expected from the hypothesis of a terrestrial origin. Gockel confirmes the conclusion of Pacini and concludes ``that a non-negligible part of the penetrating radiation is independent of the direct action of the radioactive substances in the uppermost layers of the Earth'' \cite{Goc1911}.

In spite of Pacini's conclusions, and of Wulf's and Gockel's puzzling results on the dependence of radioactivity on altitude, physicists were however reluctant to give up the hypothesis of a terrestrial origin. The situation was cleared up thanks to a long series of balloon flights by the Austrian physicist Victor Hess, who established the extra-terrestrial origin of at least part of the radiation causing the observed ionisation. 

Hess was born in 1883 in Steiermark, Austria. After graduating in Graz in 1906, he became {\em Privatdozent an der Wiener Universit\"at} in 1910.  He worked under professor Meyer at the Institute of Radium Research of the Viennese Academy of Sciences, where he performed most of his work on cosmic rays, and in 1919 he became professor of Experimental Physics at the Graz University. Hess was on leave of absence from 1921 to 1923 and worked in the United States, where he took a post as Director of the Research Laboratory of the United States Radium Corporation, at Orange (New Jersey). In 1923 he returned to Graz University and in 1931 he moved to Innsbruck as a professor. In 1936 Hess was awarded the Nobel Prize in physics for the discovery of cosmic rays. After moving to the USA in 1938 as professor at Fordham University, Hess became an American citizen in 1944, and lived in New York until his death in 1964.

Hess started his experiments by studying Wulf's results, and he knew the detailed predictions by Eve \cite{Eve1911} on the coefficients of absorption for radioactivity in the atmosphere. Eve wrote that, if one assumed a uniform distribution of RaC on the surface and in the uppermost layer of the Earth, ``an elevation to 100 m should reduce the [radiation] effect to 36 percent of the ground value''. Hess added: ``This is such a serious discrepancy [with Wulf's results] that its resolution appears to be of the highest importance for the radioactive theory of atmospheric electricity'' \cite{Hes1912}. Since in the interpretation of Wulf's and Gockel's results the absorption length of the radiation (at that time identified mostly with gamma radiation) in the air was crucial, Hess decided first to improve the experimental accuracy of Eve's result by ``direct measurements of the absorption of gamma rays in air'' \cite{Hes1911}. He used  probes of about 1 g RaCl$_2$ at distances up to 90 m, and obtained an absorption coefficient consistent with Eve. Hence the contradiction of Wulf's results remained; Hess concluded that ``a clarification can only be expected from further measurements of the penetrating radiation in balloon ascents'' \cite{Hes1911}.

\begin{figure}
\begin{center}
\resizebox{0.7\columnwidth}{!}{ \includegraphics{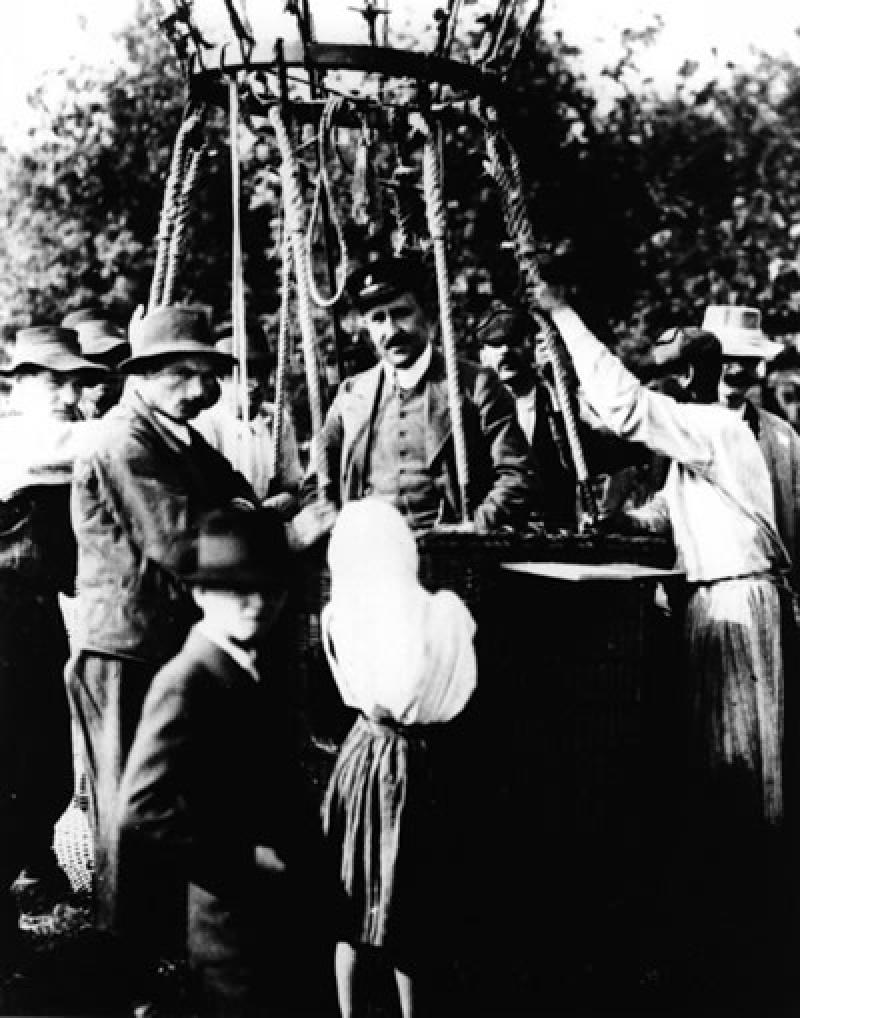} }
\end{center}
\caption{Historical photograph of Hess preparing for a balloon flight (American Physical Society).}
\label{fig:balloon}       
\end{figure}

Hess continued his studies with balloon observations (Fig. \ref{fig:balloon}). The first ascension took place on August 28, 1911. ``[T]he balloon `Radetzky' of the Austrian aeroclub É with Oberleutnant  S. Heller as pilot and me as sole passenger was lifted'' \cite{Hes1911}. The ascension lasted four hours and went up to a height of 1070 m above ground. A second ascension was done in another balloon (`Austria') during the night of 12 October 1911. During both balloon flights, the intensity of the penetrating radiation was measured to be constant with altitude within errors.

\begin{figure}
\begin{center}
\resizebox{\columnwidth}{!}{  \includegraphics{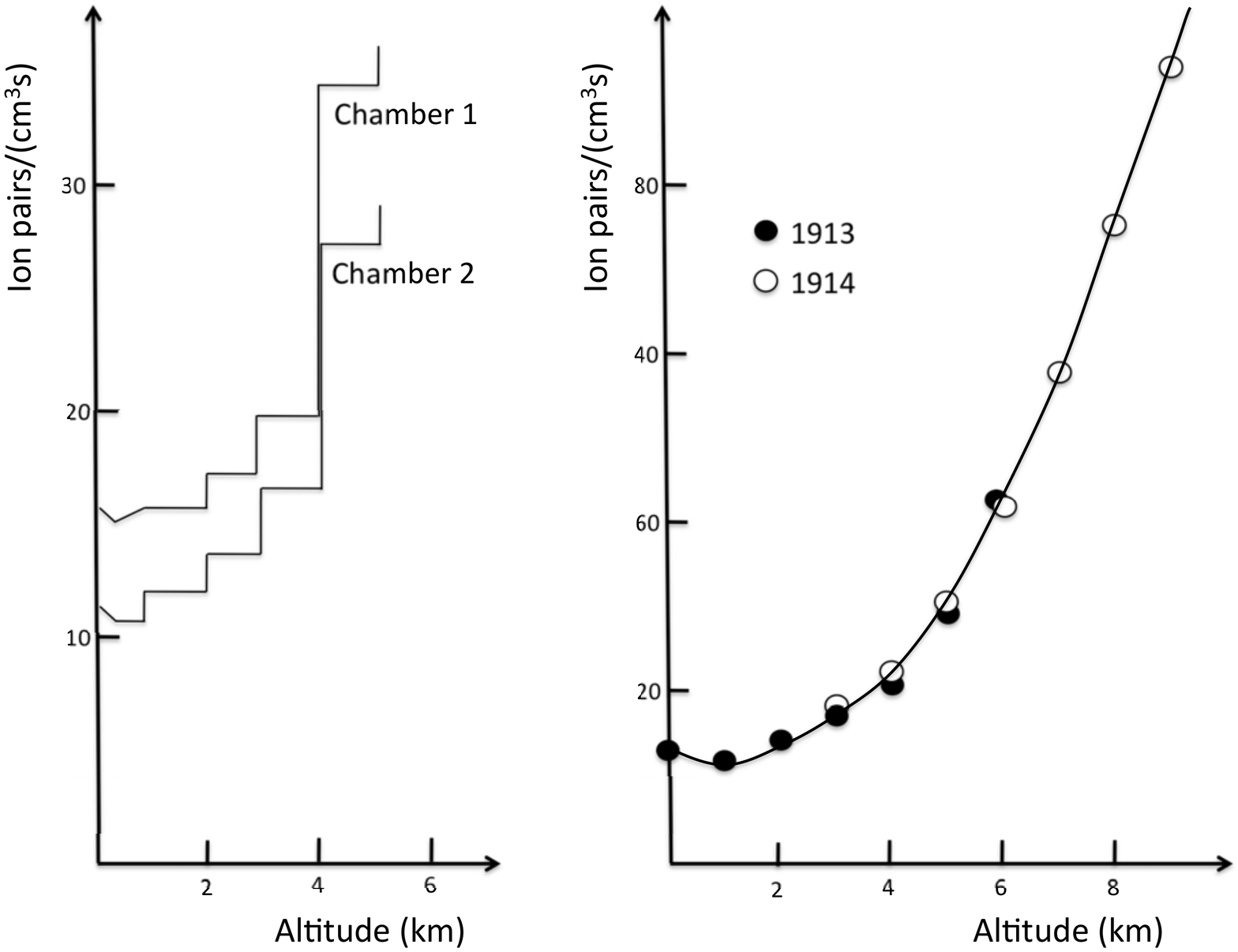} }
\end{center}
\caption{Variation of ionisation with altitude. Left panel: Final ascent by Hess (1912), carrying two ion chambers. Right panel: Ascents by Kolh\"orster (1913, 1914). Original drawing by A. De Angelis.}
\label{fig:increase}       
\end{figure}

From April 1912 to August 1912 Hess had the opportunity to fly seven times with three instruments (since for a given energy electrons have a shorter range than heavier particles, one of the three instruments had a thin wall to estimate the effect of beta radiation). In the final flight, on August 7, 1912, he reached 5200 m. To his surprise, the results clearly showed that the ionisation, after passing through a minimum, increased considerably with height (Fig. \ref{fig:increase}).  ``(i) Immediately above ground the total radiation decreases a little. (ii) At altitudes of 1000 to 2000 m there occurs again a noticeable growth of penetrating radiation. (iii) The increase reaches, at altitudes of 3000 to 4000 m, already 50\% of the total radiation observed on the ground. (iv) At 4000 to 5200 m the radiation is stronger [more than 100\%] than on the ground'' \cite{Hes1912}.

Hess concluded that the increase of the ionisation with height must be due to radiation coming from above, and he thought that this radiation was of extra-terrestrial origin. He also excluded the Sun as the direct source of this hypothetical penetrating radiation due to there being no day-night variation. Hess finally published a summary of his results in Physikalische Zeitschrift in 1913 \cite{Hes1913}, a paper which reached the wide public.

The results by Hess were later confirmed by Kolh\"orster \cite{Kol1914} in a number of flights up to 9200 m.  An increase of the ionisation up to ten times that at sea level was found. The absorption coefficient of the radiation was estimated to be 10$^{-5}$ per cm of air at NTP. This value caused great surprise as it was eight times smaller than the absorption coefficient of air for gamma rays as known at the time.

Hess coined the name H\"ohenstrahlung after the 1912 flights. Hess himself notes in a 1931 paper \cite{Hess1931} that there were several other names used  to indicate the extraterrestrial radiation before ``cosmic rays'', suggested later by Millikan, was generally accepted: Ultrastrahlung, kosmische Strahlung, Ultra-X-Strahlung.

\section{Developments after the 1914 - 1918 war}

During the war in 1914 - 1918 and the following years thereafter very few investigations of the penetrating radiation were performed.  Kolh\"orster continued his investigations using newly constructed apparatuses and made measurements at mountain altitudes with results published in 1923 in agreement with earlier balloon flights. There were, however, also negative attitudes in Europe against an extra-terrestrial radiation. Hoffmann, using newly developed electrometers  \cite{hoffa,hoffb,hoffc}, concluded \cite{hoff0a,hoff0b} that the cause of ionisation was radioactive elements in the atmosphere. Similar conclusions were reached by Behounek \cite{beh}.

After the war, the focus of research moved to the US. Millikan and Bowen \cite{Mil1923} developed a low mass (about 200 g) electrometer and ion chamber for unmanned balloon flights using data transmission technology developed during World War I. In balloon flights to 15000 m in Texas they were surprised to find radiation intensity not more than one-fourth the intensity reported by Hess and Kolh\"orster. They attributed this difference to a turnover in the intensity at higher altitude, being unaware that a geomagnetic effect existed between the measurement in Europe and Texas. Thus, Millikan believed that there was no extraterrestrial radiation. As reported to the American Physical Society in 1925 Millikan's statement was ``The whole of the penetrating radiation is of local origin''.

In 1926, however, Millikan and Cameron \cite{Cam1926} carried out absorption measurements of the radiation at various depths in lakes at high altitudes. Based upon the absorption coefficients and altitude dependence of the radiation, they concluded that the radiation was high energy gamma rays and that ``these rays shoot through space equally in all directions'' calling them ``cosmic rays''. 

Millikan was handling with energy and skill the communication with the media, and in the US the discovery of cosmic rays became, according to the public opinion, a success of American science. Millikan argued that radiations are ``generated by nuclear changes having energy values not far from [those that they recorded] in nebulous matter in space.'' Millikan then proclaimed that this cosmic radiation was the ``birth cries of atoms'' in our galaxy. His lectures drew considerable attention from, among others, Eddington and Jeans, who struggled unsuccessfully to describe processes that could account for Millikan's claim \cite{simpson}.

\subsection{Properties of the radiation}
It was generally believed that cosmic radiation was gamma radiation because of its penetrating power (one should remember that the penetrating power of relativistic charged particles were not known at the time). Millikan had put forward the hypothesis that gamma rays were produced when protons and electrons form helium nuclei in the interstellar space. 

A key experiment, which would decide the nature of cosmic rays (and in particular if they were charged or neutral), was the measurement of the dependence of cosmic ray intensity on geomagnetic latitude. Important measurements were made in 1927 and 1928 by Clay \cite{clay1,clay2} who, during two voyages between Java and Genova, found that ionisation increased with latitude. No such variation was expected if the radiation was gamma radiation, but Clay did not draw a firm conclusion as to the nature of cosmic radiation. Clay's work was disputed by Millikan.

With the introduction of the Geiger-Muller counter in 1928, a new era began and soon confirmation came that the cosmic radiation is indeed corpuscular. Kolh\"orster introduced the coincidence technique. Bothe and Kolh\"orster \cite{BK} concluded that cosmic radiation is mainly or fully corpuscular, but still Millikan did not accept this view. 

In 1932 Compton carried out a world-wide survey to settle the dispute.  He reported \cite{Com1933a,Com1933b} that there was a latitude effect, that cosmic rays were charged particles and that Millikan was wrong.  Millikan strongly attacked Compton, but after repeating his experiment in 1933 he admitted that there was a latitude effect and that the cosmic rays must be (mostly) charged particles. In 1933, three independent experiments by
Alvarez \& Compton \cite{aecew}, Johnson \cite{jew}, Rossi \cite{rew} discovered that more cosmic rays were coming from West than from East close to the Equator: this is due to the interaction with the magnetic field of the Earth, and it demonstrated that cosmic rays are mostly positively charged particles. However, it would take until 1941 before it was established in an experiment by Schein \cite{sche} that cosmic rays were mostly protons.

Much attention in the middle of the 1920's was given to the question of variation  of radiation with time. Many investigators had found such variations, but towards 1930 the general opinion was that no such variation existed. The work of Forbush led to the proof that the observed intensity of cosmic rays within the atmosphere of Earth varied with time \cite{Forbush1938}. 

\subsection{Cosmic rays and the beginning of particle physics}
Thanks to the development of cosmic ray physics, scientists knew then that astrophysical sources were providing very-high energy bullets entering the atmosphere. It was then obvious to investigate the nature of such bullets, and to use them as probes to investigate matter in detail, such as along the lines of the experiment made by Rutherford in 1900. Particle physics, the science of the fundamental constituents of matter, started with cosmic rays. 
Many fundamental discoveries were made. 

The first was the discovery of the positron. While watching the tracks of cosmic rays passing through his cloud chamber, Anderson in 1933 discovered antimatter in the form of the anti-electron, later called the positron \cite{positron}. 

Neddermeyer and Anderson discovered in 1937 in cosmic rays the elementary subatomic particle called the muon \cite{muon}. The positron and the muon were the first series of subatomic particles discovered using cosmic rays. Later came the discovery of the first meson, the charged pion by Powell, Lattes and Occhialini in 1947 \cite{pion}. The discovery of strangeness \cite{kaon} was also made thanks to cosmic rays. Particle physicists used cosmic rays as the main tool for their research, and the pioneering results of that field are due to cosmic rays.  This includes the beginning of the discovery of new particles
(the so-called ``particle zoo'') in the tracks 
of atmospheric emulsions and chambers, and was the motivation of unifying and simplifying ideas like the theory of quarks. A historical account is 
provided, for example in \cite{birth}.

World War II interrupted the study of high-energy physics through cosmic rays but provided new resources, both technical and political, for the study of elementary particles. Technical resources included advances in microwave electronics and the design of human-made particle accelerators, which allowed physicists to produce high energy particles in a controlled environment. By about 1950 elementary particle physics would have been dominated by accelerator physics, at least until the beginning of the 1990s when possible exploration with the energies one can produce on Earth started showing signs of saturation.

Before and during the World War II Werner Heisenberg, at that time in Leipzig, published several papers on cosmic rays. In particular the shower phenomenon was the subject for his interest (see, for axample \cite{Heisenberg1936,Heisenberg1938}). In 1943, during the war, Heisenberg edited a Festschrift about cosmic rays for Sommerfeld \cite{Heisenberg1943}. This work includes theoretical papers from presentations at the Kaiser-Wilhelm institute in Berlin in 1941-42; five of the papers are by Heisenberg himself. In the preface, dated in June 1943, Heisenberg describes the difficult situation for science in Germany. He points at the lack of information from foreign countries and the absence of reports and publications from the US from the summer of 1941. The situation is similar to the one during and immediately after World War I (see section 5.2).

\subsection{Cosmic-ray physics today}
Cosmic rays are today central to the new field of astroparticle physics, an interdisciplinary field between astrophysics, cosmology and particle physics. Astroparticle physics has grown considerably in the last 20 years \cite{cirkel} and many large projects are being performed, searching for, for example the dark matter in the Universe. 

Satellite projects like the Fermi gamma-ray space telescope\footnote{Web site http://fermi.gsfc.nasa.gov/}
and the PAMELA magnetic spectrometer\footnote{Web site http://pamela.roma2.infn.it} have given new cutting-edge results; PAMELA in particular \cite{pamela2} observed a yet unexplained anomalous yield of positrons, which might point to new physics, in particular related to dark matter. The planned AMS-02  mission\footnote{Web site http://ams.cern.ch/} will further extend this type of research on the presence of antimatter in the Universe. 

The study of very highest energy cosmic-ray showers, 70 years after the discovery of air showers by Rossi and Auger, is providing the large ground-based Pierre Auger Observatory\footnote{Web site http://www.auger.org/}  fundamental knowledge of the spectrum and possible other sources \cite{augsc1,augsc2} of cosmic rays. In particular the region near the GZK cutoff is explored\footnote{A theoretical upper limit on the energy of cosmic rays from distant sources  was computed in 1966 by Greisen~\cite{gzk1}, Kuzmin and Zatsepin \cite{gzk2}, and it is called today the GZK cutoff.
	Protons with energies above a threshold of about 10$^{20}$ eV suffer a resonant interaction with the cosmic microwave background photons to produce pions ($p$+$\gamma$$\rightarrow$$\Delta$$\rightarrow$$N$$+$$\pi$).
	This continues until their energy falls below the production threshold.
	Because of the mean path associated with the interaction, extragalactic cosmic rays from distances larger than 50 Mpc from the Earth and with energies greater than this threshold energy should be strongly suppressed on Earth, and there are no known sources within this distance that could produce them.}; anomalies in the cosmic rays yield near the GZK cutoff might point to new physics, in particular to violations of the Lorentz invariance at very high energy, as envisaged, for example in some Quantum-Gravity inspired scenarios \cite{ameli}.

The ground-based very-high energy gamma telescope arrays H.E.S.S.\footnote{Web site http://www.mpi-hd.mpg.de/hfm/HESS/}
MAGIC\footnote{Web site http://wwwmagic.mppmu.mpg.de ; http://magic.fisica.uniud.it} and 
VERITAS\footnote{Web site http://veritas.sao.arizona.edu/} are mapping the cosmic emitters of gamma rays (and thus, indirectly, of cosmic rays) in the TeV region; together with Fermi, they are providing indications of a link between the photon accelerators and the accelerators of cosmic rays in the Milky Way, supernova remnants in particular \cite{zwi}. Studying the propagation of very-high-energy photons through cosmological distances, they are also sensitive to possible violations of the Lorentz invariance at very high energy levels \cite{ameli}.

In view of the fundamental limitations of earthbound high-energy physics experiments, cosmic rays and cosmological sources again move into the focus of very-high-energy particle and gravitational physics. 


\section{The Nobel Prize to Hess in 1936}
\begin{figure}
\begin{center}
\resizebox{0.7\columnwidth}{!}{%
\includegraphics{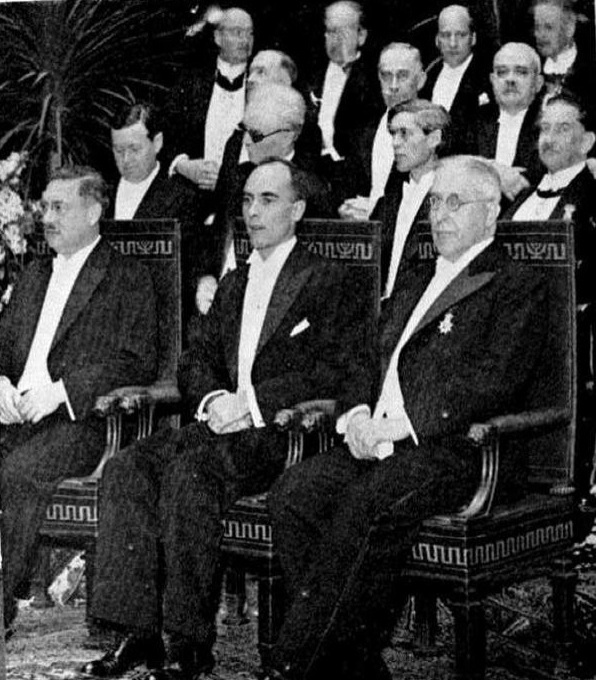} }
\end{center}
\caption{Waiting to receive their prizes on 10 December 1936, from left Petrus Debye (chemistry winner), Carl Anderson and Victor Hess. Behind them previous laureates, including Manne Siegbahn (physics 1924), Gustav Dal\'en (physics 1912) and The Svedberg (chemistry 1926). Courtesy of the Nobel foundation.}
\label{fig:hessnobel}       
\end{figure}

The  1936 Nobel Prize in Physics was shared by Professor V.F. Hess for the discovery of cosmic rays  and Dr. C.D. Anderson for the discovery of the positron (Fig. \ref{fig:hessnobel}).

The Royal Swedish Academy of Sciences had received a total of 22 Prize proposals from 31 nominators for 18 different Prize Candidates in 
Physics\footnote{The reports and other information relative to the 1936 Nobel Prize in Physics referred to in the text is unpublished material, but available for scientific historical research at the Center for History of Science at the Royal Swedish Academy of Sciences in Stockholm, Sweden.}. Hess was nominated by J. Clay, Amsterdam, for a non-shared prize and by A.H. Compton, Chicago, for a prize shared with J. Clay, Amsterdam. Compton also nominated C.D. Anderson for the discovery of the positron. Hess had been nominated the first time in 1931 by Pohl, from G\"ottingen, and then in 1933 by Plotnikov, from Zagreb and in 1934 by Willst\"atter, from Munich.
 
In its recommendation to the Academy, the Committee points out that the discovery of cosmic rays has opened new areas for experimental and theoretical physics of greatest significance to our understanding of the structure and origin of matter. It is clear, the Committee says, that Hess with his skillful experiments has proven the existence of an extraterrestrial penetrating radiation, a discovery more fundamental than that of the radiation's corpuscular nature and that of the latitude variation of its intensity.

In his detailed nomination Compton had written: ``The time has now arrived, it seems to me, when we can say that the so-called cosmic rays definitely have their origin at such remote distances from the Earth that they may properly be called cosmic, and that the use of the rays has by now led to results of such importance that they may be considered a discovery of the first magnitude. ... It is, I believe, correct to say that Hess was the first to establish the increase of the ionisation observed in electroscopes with increasing altitude; and he was certainly the first to ascribe with confidence this increased ionisation to radiation coming from outside the Earth''. Compton points out the importance of the discovery for other areas of physics and writes: ``Before it was appropriate to award the Nobel Prize for the discovery of these rays, it was necessary to await more positive evidence regarding their unique characteristics and importance in various fields of physics. This has now been accomplished. Studies of the magnetic latitude effect on cosmic rays have shown that they include electrical particles of much higher energy than are available from artificial sources, further that these rays come from a source which may be properly called cosmic. The usefulness of the rays has been demonstrated by the experiment which have revealed the existence of the positron.''

The nine-page report on cosmic rays, commissioned by the Committee, was prepared by E. Hulth\'en, one of the members of the Committee (other members in 1936 were H. Pleijel, chair, A.E. Lindh, C.W. Oseen and M. Siegbahn). It was submitted to the Committee in June 1936 and was included as an appendix to the Committee's proposal to the Academy.

Hulth\'en notes that as regards to the nomination of Hess, the discovery is quite old. However, the nominators for the 1936 prize both point out that the cosmic origin of the penetrating radiation only recently has been fully established and that it has been very important for other discoveries. In his careful and complete report Hulth\'en introduces the work of Hess with the measurements of the remaining ionisation observed inside electroscopes 
 by Wilson  and by Elster and Geitel. Further research by Rutherford and Cooke, by Mc Lennan and Burton showed that the observed ionisation was reduced when the electroscope was surrounded by absorbers. Hulth\'en points out that at the time the research on radioactivity, discovered in 1896 by Becquerel, was in its infancy and radioactive substances could therefore be a natural explanation of the origin of the remaining ionisation. Based on experiments by Eve, Pacini, Mache and others it became clear that only a small fraction of the remaining ionisation could be attributed to radioactive substances. Based on estimates of the absorption of gamma rays estimates were made on the ionisation at different altitudes. However, early balloon based measurements by Bergwitz and Gockel  did not show a very significant decrease of the ionisation. 

Hulth\'en notes - and quotes - Gockel's conclusion that the results of his balloon measurements, in agreement with measurements of Pacini, show that a not insignificant part of the radiation is independent of direct action of substances in the crust of the Earth.
He notes however that Hess' careful work includes an accurate measurement of the absorption of gamma rays, confirming the results of Eve, and several balloon ascents in 1911 and 1912, finally finding an increase by a factor of two in the ionisation at an altitude of 5200 m. Hulth\'en quotes Hess' conclusion that the results show that a very penetrating radiation is incident on the atmosphere from the outside (`ÉDie Ergebnisse der vorliegenden Beobachtungen scheinen am ehesten durch die Annahme erkl\"art werden zu k\"onnen, dass eine Strahlung von sehr hoher Durchdringungskraft von oben her in unsere AtmosphŠre eindringt...' \cite{Hes1913}). Hess'  results attracted a lot of attention, but were also questioned for uncertain experimental arrangements. Very soon, however, confirming results came  from Kolh\"orster, who found an ionisation 40 times larger at an altitude of 9300 m.  
After the World War I research was resumed in the 1920s. Hess'  and Kolh\"orster's results were at first still questioned by among others Millikan, who, however, in 1925 confirmed Hess' conclusion.

Hulth\'en's report concludes with a discussion of the importance of Hess' results for other areas of fundamental physics.

The physics class of the Academy met on  October 29th,  1936, and the committee proposal was seconded. The Academy met 12 November 1936 and the 78 present members decided to follow the proposal from the committee and awarded the 1936 Nobel Prize in Physics to V.F. Hess for the discovery of cosmic rays and to C.D. Anderson for the discovery of the positron.

\section{Discussion}

Hess is today remembered as the discoverer of cosmic rays for which he was awarded the 1936 Nobel Prize in physics; his discovery was based on contributions of many other scientists. It seems to us that in particular the important contribution by Pacini has been forgotten. Pacini reached important conclusions on the origin of the ``penetrating radiation'' one year before Hess; the technique used by Pacini, however, could not disprove a possible atmospheric origin of
the background radiation. Pacini's work is not often cited in reviews of the history of cosmic rays. It was however correctly cited in the report of the Nobel Committee in 1936. Pacini was never nominated for the Nobel Prize. In fact of the reviews cited  \cite{montgo,janossy,leprince,rossi,hillas,wilson,wigandr,millikanr,comptonr,swann,xu,ginz,puppi} only J\'anossy \cite{janossy} and Wigand \cite{wigandr} mention the work of Pacini. 

Scientific research is today characterised by openness and rapid communication of results. This was not the case when cosmic rays were discovered. Communication was slow, there were language barriers combined with nationalism and there were important effects of World War I. Several other causes might have contributed to the lack of credit given in present days to Pacini's work.

\subsection{Recognition of Pacini's work}

The first claims by Pacini \cite{Pac1909} came before any work by Hess. However in his
definitive paper \cite{Pac1912}, published a few months before the paper by Hess, Pacini was
evidently aware of the preliminary (and non conclusive) results obtained by Hess
in 1911 as he did quote them correctly.\footnote{In the fundamental paper in which he publishes for the first time a significant effect \cite{Hes1912}, Hess quotes the partial 1908-1910 results by Pacini \cite{Pac1910b}, but not his conclusive (statistically significant) results in 1911 \cite{Pac1912}. Unfortunately we do not know the times of submission of the two articles by Hess and Pacini; slowness of communication is a likely explanation. However, in the subsequent article by Hess, summarising his measurements and the hypotheses on the origin of radiation \cite{Hes1913}, Pacini's works is not cited.} Some excerpts from mail exchanges\footnote{The text of the first three letters reported in this section has been quoted in the commemoration by G.B. Rizzo at the University of Bari in July 1934 \cite{Riz1934b}, where Hess' letters were translated from German into Italian;  a transcription of part of the letters by Hess, in German, is reported in \cite{bologna}, together with Hess' letter dated May 20 quoted in this Section.  The obituary by Professor Rizzo says that all papers, books and letters belonging to Pacini have been donated by the widow to the University of Bari; unfortunately the Pacini archive has not been found at the University.}
 that occurred
between the two scientists in 1920 \cite{Riz1934b,bologna} are very illuminating. 

On March 6, 1920, Pacini wrote to Hess: ``...I had the opportunity to study some of your papers about electrical-atmospherical phenomena that you submitted to the Principal Director of
the Central Bureau of Meteorology and Geodynamics [in Rome]. I was already aware of some of these works from summaries that had been reported to me during the war. [But] the paper \cite{Hess1919} entitled  `The problem of penetrating radiation of extraterrestrial origin'\footnote{`Die Frage der durchdringenden Strahlung ausserterrestrischen Ursprunges'} was unknown to
me. While I have to congratulate you for the clarity in which this important matter is
explained, I have to remark, unfortunately, that the Italian measurements and observations,
which take priority as far as the conclusions that you, Gockel and Kolh\"orster
draw, are missing; and I am so sorry about this, because in my own publications I
never forgot to mention and cite anyone...''.

The answer by Hess, dated March 17, 1920, was: ``Dear Mr. Professor, your very
valuable letter dated March 6 was to me particularly precious because it gave me the
opportunity to re-establish our links that unfortunately were severed during the war.
I could have contacted you before, but unfortunately I did not know your address. My
short paper `The problem of penetrating radiation of extraterrestrial origin'
is a report of a public conference, and therefore has no claim of completeness. Since
it reported the first balloon measurements, I did not provide an in-depth explanation
of your sea measurements, which are well known to me. Therefore please excuse me
for my unkind omission, that was truly far from my aim ...''. 
On April 12, 1920,
Pacini in turn replied to Hess: ``... [W]hat you say about the measurements on the
penetrating radiation performed on balloon is correct; however the paper `The problem of penetrating radiation of extraterrestrial origin'  lingers quite a bit on
measurements of the attenuation of this radiation made before your balloon flights,
and several authors are cited whereas I do not see any reference to my relevant
measurements (on the same matter) performed underwater in the sea and in the
Bracciano Lake, that led me to the same conclusions that the balloon flights have
later confirmed.''

Finally, on May 20, 1920, Hess replied to Pacini: ``...Coming back to your publication in `Nuovo Cimento', (6) 3 Vol. 93, February 1912, I am ready to acknowledge that certainly you had the priority in expressing the statement, that a non terrestrial radiation of 2 ions/cm$^3$ per second at sea level is present.  However, the demonstration of
the existence of a new
source of penetrating radiation
from above
came from my balloon ascent to a height of 5000 meters on August 7 1912, in which I have discovered a huge increase in radiation above 3000 meters.''

The Hess-Pacini correspondence, nine years after Pacini's work and eight years after Hess' 1912 balloon flight, shows how difficult communication was at the time. Also language difficulties may have contributed: Pacini publishing mostly in Italian and Hess in German.

\begin{figure}
\begin{center}
\resizebox{0.98\columnwidth}{!}{%
\includegraphics{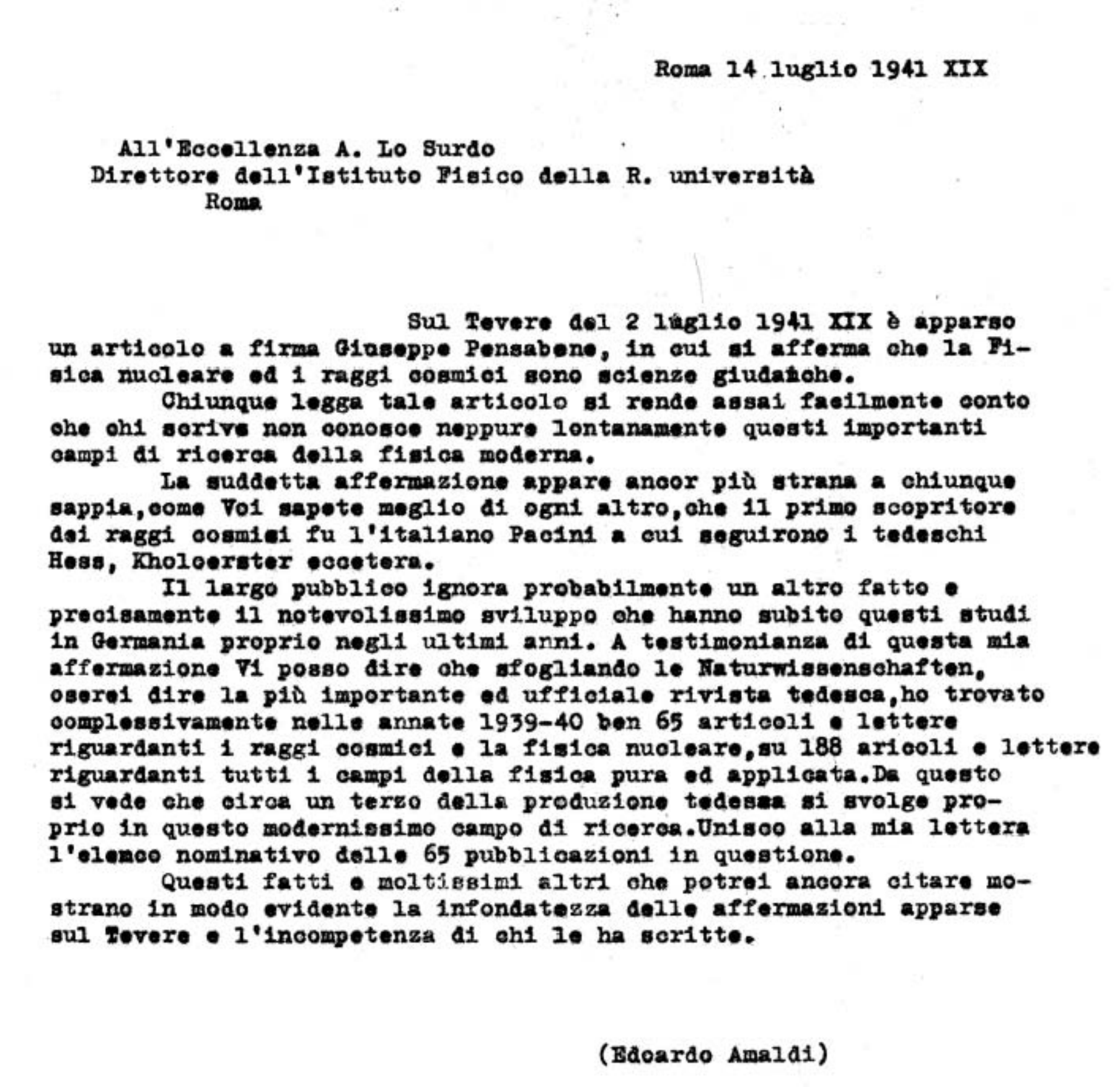} }
\end{center}
\caption{Letter by E. Amaldi to Lo Surdo. The letter belongs to the Amaldi Archive (scatola 212, fasc. 1) at the ``La Sapienza'' University of Rome (courtesy of U. Amaldi and G. Battimelli).}
\label{fig:amaldilet}       
\end{figure}

Edoardo Amaldi\footnote{Edoardo Amaldi (Carpaneto Piacentino, 1908 - Rome, 1989) worked closely with Enrico Fermi (he had been  one of the ``boys from Via Panisperna'') until 1938, when Fermi 
was forced to leave Italy. Professor of experimental physics in Rome for more than 40 years, he was co-founder of CERN, of which he was Secretary-General (this was the name of what is called today Director-General) during the years 1952-1954, of ESA, and of the Italian National Institute for Nuclear Physics (INFN), of which he was President during the years 1960-1965.
His fields of interest were elementary particle physics, nuclear physics, and the physics of cosmic radiation, with a particular attention to the study of gravitational waves \cite{ama1,trec}.} 
 had however no doubt that Domenico Pacini was indeed the discoverer 
of cosmic rays as stated in a letter (Fig. \ref{fig:amaldilet})  that he wrote on July 14, 1941 to the director of the Physics Institute of Rome, Antonino 
Lo Surdo\footnote{Antonino Lo Surdo (Siracusa, 1880 - Rome,  1949) was professor of physics in the University of Rome since 1919; he became director of the Physics Institute in 1937 \cite{trec}.}.

E. Amaldi's letter was motivated by an article that had appeared on July 2, 1941 in a local newspaper, Il Tevere, in which it was stated that nuclear physics and cosmic ray physics were ``Jewish sciences''.

E. Amaldi writes: ``... this statement appears so strange to anyone who
knows, as you certainly do, that the Italian Domenico Pacini [(a non-Jew)] was the discoverer of
the cosmic rays ...''

Pacini's work was carried out in difficult conditions because of lack of resources
available to him and because of lack of scientific freedom (during the crucial years he
was working at the Central Bureau of Meteorology and Geodynamics and was unable to attend international conferences). It is also important to realise that his work was met with substantial indifference by the Italian academic world that, per se, made Nobel Prize nominations difficult. 

\subsection{Nationalism}
At the turn of the century, coinciding with the first results from electroscopes,
the first Nobel prizes were delivered. The prize ceremonies in Stockholm became celebrations
of scientific internationalism. However, with the outbreak of World War I, disturbances and nationalism grew in international scientific relations. 


This has been studied by many authors and there is a rich literature on nationalism and social context of physics. A good bibliography can be found in for example \cite{birth}. We refer here to the study of nationalism and internationalism in science 1880 -- 1939 by Crawford \cite{craw}.



Crawford uses the Nobel population and shows that, e.g., in physics and chemistry, the proportions of nominations received by ``Central Power" candidates (Central power countries included Germany and Austria, whereas Allied countries included France, Italy and the United Kingdom) from nominators in Allied countries dropped to almost zero in the period 1916-1920. As an example Planck was nominated for the Nobel Prize 20 times in the period from 1916 to 1920, 18 of which were from Germany.

Before the war Germany was the leading industrial nation and produced in 1913
as much goods as France and UK combined; Austria enjoyed an exceptional cultural
time for art, architecture and literature. As the war approached, in Germany propaganda
manifestoes were signed that put forward German scientists. The ``Appeal to the
Cultured World'', a October 1914 manifesto, denying German responsibility for the war, was signed by 93 intellectuals including several scientists and Nobel laureates, among them R\"ontgen and Planck. In 1915 a manifesto was signed by several physicists that declared that scientific relations with England would not be resumed for the foreseeable future.

After the war the allied countries set up the International Research Council (IRC)
in 1919 \cite{IRC}. The IRC practiced an exclusionary policy with respect to the former
Central Power and there was at the time a widespread opinion that Germany should
be chastised. Hence, in the early post-war period German scholars could or would
not attend international meetings or otherwise participate in international scientific
activities. A breakthrough occurred in 1926 when the exclusion of German scientists
dropped to about 15\% of the meetings, a normalisation of international scientific
relations. This happened at about the same time as the extraterrestrial origin of the
cosmic radiation was generally accepted.

It cannot be excluded that the dispute between Millikan in the US and Hess and others in Europe, as illustrated by the 1928 paper by Bergwitz, Hess et al. \cite{BergwitzHess}, also has a nationalistic side; the authors there claim they are ``establish[ing] the history of the discovery of cosmic rays''.
A deep analysis of the late recognition of the discovery of cosmic rays, and of its possible relation to rivalries between Europe and the US,  has been done in \cite{demaria}.

\section{Conclusion}

The work behind the discovery of cosmic rays, a milestone in science, comprised scientists in Europe
and the New World and took place during a period characterised by lack of communication and by nationalism caused primarily by the World War I. The many scientists that took part in this research starting a century ago, either alone or as a two-person group, were fascinated by the penetrating radiation and wanted to understand the origin and properties of it. It took from the turn of the century until 1926 before the extraterrestrial nature of the penetrating radiation was generally accepted.

In the work that culminated with high altitude balloon flights, many important contributions have been forgotten and in particular those of Pacini in 1909-1911. Several historical, political and personal facts might have contributed to the lack of references to the work of Pacini in the history of cosmic rays. 

The nominations for a Nobel Prize in physics for the discovery of cosmic rays started only in the 1930s.  Pacini was never nominated and died two years before the Nobel Prize was given to Hess in 1936.

%

\section*{Acknowledgements}

We are grateful to the University of Bari, and in particular to Professor A.~Garuccio, 
for supporting the research of documents regarding Domenico Pacini; and  to the Dipartimento 
Interateneo di Fisica of Bari for organizing the Domenico Pacini memorial day that 
was held in Bari on April 17, 2007.

Professors F.~Guerra and N.~Robotti uncovered the letter by E. Amaldi to ``Il Tevere'' and the dossier of the applications by Pacini for the ``libera docenza'' and for the professorship in the Amaldi Archive at Rome's ``La Sapienza'' University and in the Biblioteca Nazionale; R. Garra uncovered material in ``Collegio Romano''.  We thank Professors U. Amaldi and G. Battimelli for allowing us to reproduce the letter by E. Amaldi.

 Comandante E.~Bagnasco and A.~Lombardi from Associazione Culturale Italia, C.~D'Adamo and H. von Zeschau from 
 the Associazione Regia Marina Italiana, provided historical pictures and information on the ships by the Italian Navy.
 
 Professors A. Bettini,  L.~Cifarelli, B. De Lotto, L.~Guerriero, E.~Menichetti, M.~Persic,  R. Paoletti, P.~Spinelli,  and Dr. P.~Lipari provided help, support, discussions and suggestions.   Professors H. Rechenberg and  A.A. Watson provided precious historical information.

The detailed study of the documents relative to the Nobel Prize to Hess in 1936 was done at the Royal Swedish Academy of SciencesÕ Center for History of Science. The help and support  provided by prof. K. Grandin of the Center is greatly acknowledged. 

One of us (AdA) would like to thank the Max Planck Institute in Munich, and Professor M. Teshima, Dr. R. Mirzoyan and the MAGIC MPI group, for the kind hospitality in Germany.

Finally, a special acknowledgement goes to Professors  N. Giglietto and S. Stramaglia from Bari, who uncovered a lot of material, and transmitted their enthusiasm for giving an objective account of the contribution by Domenico Pacini to the discovery of cosmic rays.

\end{document}